\documentclass[usenatbib,useAMS,fleqn]{mn2e}

\newif\ifpdf\ifx\pdfoutput\undefined\pdffalse\else\pdfoutput=1\pdftrue\fi

\usepackage{epsf}

\usepackage{graphics}
\usepackage{amssymb}
\usepackage{amsmath}

\newcommand{\ee}{\ensuremath{{\rm e}}}
\newcommand{\ii}{\ensuremath{{\rm i}}}

\newcommand{\diff}{\ensuremath{{\rm d}}}

\newcommand{\Lstar}{\ensuremath{L_{\ast}}}
\newcommand{\Rstar}{\ensuremath{R_{\ast}}}
\newcommand{\Mstar}{\ensuremath{M_{\ast}}}

\newcommand{\Teff}{\ensuremath{T_{\rm eff}}}

\newcommand{\Lrad}{\ensuremath{L_{\rm R}}}
\newcommand{\lrad}{\ensuremath{l_{\rm R}}}

\newcommand{\veq}{\ensuremath{v_{\rm e}}}
\newcommand{\vsini}{\ensuremath{v\sin i}}
\newcommand{\vcrit}{\ensuremath{v_{\rm c}}}

\newcommand{\ncowl}{\ensuremath{\tilde{n}}}

\newcommand{\nPml}{\ensuremath{\bar{P}^{m}_{\ell}}}
\newcommand{\hough}{\ensuremath{\Theta^{m}_{\ell}}}
\newcommand{\lamml}{\ensuremath{\lambda^{m}_{\ell}}}
\newcommand{\Cnl}{\ensuremath{C_{\ncowl,\ell}}}
\newcommand{\elle}{\ensuremath{\tilde{\ell}}}

\newcommand{\pfreq}{\ensuremath{\sigma}}
\newcommand{\pfreqr}{\ensuremath{\sigma_{\rm r}}}
\newcommand{\pfreqi}{\ensuremath{\sigma_{\rm i}}}
\newcommand{\dfreq}{\ensuremath{\omega}}
\newcommand{\pper}{\ensuremath{P}}
\newcommand{\pperi}{\ensuremath{P_{\rm i}}}

\newcommand{\rfreq}{\ensuremath{\Omega}}
\newcommand{\rfreqv}{\ensuremath{\bmath{\Omega}}}
\newcommand{\rfreqc}{\ensuremath{\Omega_{\rm c}}}
\newcommand{\rper}{\ensuremath{\Pi}}

\newcommand{\brunt}{\ensuremath{\mathcal{N}}}

\newcommand{\nablad}{\ensuremath{\nabla_{\rm ad}}}
\newcommand{\epsad}{\ensuremath{\varepsilon_{\rm ad}}}
\newcommand{\epss}{\ensuremath{\varepsilon_{\rm S}}}
\newcommand{\epsn}{\ensuremath{\varepsilon_{\rm N}}}

\newcommand{\ceps}{\ensuremath{\bar{c}_{3}}}
\newcommand{\cthr}{\ensuremath{\bar{c}_{4}}}

\newcommand{\xir}{\ensuremath{\xi_{r}}}

\newcommand{\disc}{\ensuremath{D}}

\newcommand{\tauth}{\ensuremath{\tau_{\rm th}}}

\newcommand{\kms}{\ensuremath{{\rm km\,s^{-1}}}}
\newcommand{\cmss}{\ensuremath{{\rm cm\,s^{-2}}}}

\newcommand{\Lsun}{\ensuremath{{\rm L}_{\sun}}}
\newcommand{\Rsun}{\ensuremath{{\rm R}_{\sun}}}
\newcommand{\Msun}{\ensuremath{{\rm M}_{\sun}}}

\newcommand{\Kelv}{\ensuremath{{\rm K}}}

\newcommand{\days}{\ensuremath{{\rm days}}}

\newcommand{\eg}{e.g.}

\newcommand{\cf}{c.f.}
\newcommand{\ca}{ca.}

\newcommand{\boojum}{\textsc{boojum}}

\newcommand{\BV}{Brunt-V\"{a}i\"{a}s\"{a}la}

\newcommand{\noop}[1]{}

\title[Influence of the Coriolis force on slowly pulsating B Stars]%
      {Influence of the Coriolis force on the instability of
       slowly pulsating B stars}
\author[R.H.D.Townsend]
       {R. H. D. Townsend$^{1,2}$%
        \thanks{Email: rhdt@bartol.udel.edu}\\
        $^{1}$ Bartol Research Institute,
        University of Delaware,
        Newark, DE 19716, USA\\
        $^{2}$ Department of Physics \& Astronomy, 
        University College London, 
        Gower Street, London WC1E 6BT}
\date{%
Received: .................................... 
Accepted: ....................................
}

\pagerange{\pageref{firstpage}--\pageref{lastpage}}
\pubyear{2003}

\begin{document}


\maketitle

\label{firstpage}

\begin{abstract}
This paper explores the effect of rotation on the $\kappa$-mechanism
instability of slowly pulsating B stars. A new nonadiabatic code, that
adopts the so-called `traditional approximation' to treat the Coriolis
force, is used to investigate the influence exerted by rotation over
the stability of stellar models covering the mass range $2.5\,\Msun
\leq \Mstar \leq 13.0\,\Msun$. The principal finding is that, for all
modes considered apart from the prograde sectoral class, rotation
shifts the $\kappa$-mechanism instability toward higher luminosities
and effective temperatures; these shifts are accompanied by
broadenings in the extent of instability strips. Such behaviour is
traced to the shortening of mode periods under the action of the
Coriolis force. Instability strips associated with prograde sectoral
modes behave rather differently, being shifted to marginally lower
luminosities and effective temperatures under the influence of
rotation.

The implications of these results are discussed in the context of the
observational scarcity of pulsation in B-type stars having significant
rotation; various scenarios are explored to explain the apparent
dichotomy between theory and observations. Furthermore, the possible
significance of the findings to Be stars is briefly examined.
\end{abstract}

\begin{keywords}
stars: oscillation -- stars: rotation -- stars: variables: other --
stars: early-type -- stars: emission-line, Be -- methods: numerical
\end{keywords}


\section{Introduction} \label{sec:intro}

The slowly pulsating B (SPB) stars are a homogeneous class of mid-B
type, main-sequence objects that exhibit multiperiodic light and
line-profile variations over timescales on the order of 1--5
days. \citet{Wae1991} was the first to classify these stars as a
distinct group of early-type nonradial pulsators, by combining under
the same aegis the photometric variables discovered by
\citet{WaeRuf1985}, and the 53 Per spectroscopic variables first
observed by \citet{Smi1977}. The theoretical pulsation characteristics
of these objects have since been studied extensively, resulting in a
canonical picture \citep[see, \eg,][and references therein]{Pam1999}
of high-order g-mode pulsation, driven by the same iron-group $\kappa$
mechanism responsible for the instability of the $\beta$ Cep pulsators
\citep{DziPam1993}.

One outstanding issue in the understanding of SPB stars is the effect
rotation has on their instability. Photometric surveys of open
clusters by \citet{Bal1994}, \citet{BalKoe1994} and
\citet{BalLan1995,BalLan1996} failed to find any evidence for SPB-like
variability, but the authors did note that the stars observed were
characterized by moderate or rapid projected equatorial rotation
velocities ($\vsini \gtrsim 100\,\kms$). This result has led some to
suggest that rotation suppresses the $\kappa$-mechanism excitation of
g modes \citep[see, \eg,][]{Bal2001}. However, an alternative
interpretation has been explored by \citet{Tow2003b}, who examined the
effect of the rotation-originated Coriolis force on the flux
perturbations produced by low-frequency g modes. He demonstrated that
the confinement of these modes within an equatorial waveguide
\citep[\eg,][]{LeeSai1990,Tow2003a} reduces the amplitude of the
resulting photometric variations, possibly to below the sensitivity of
the observations.

This paper addresses the SPB-rotation question from a more-theoretical
perspective, by examining the effect the Coriolis force has on the
$\kappa$-mechanism instability of SPB stars. The high-order g modes
excited in these objects are characterized by low frequencies, and
their dynamics can thus be expected to be influenced appreciably by
the Coriolis force \citep[see][]{Tow2005}. For this reason, canonical
approaches that treat the rotation as a small perturbation to the
pulsation \citep[\eg,][their \S19 and references
therein]{CarHan1982,Unn1989} are not well suited to SPB
stars. Building on a methodology originated by
\citet{LeeSai1987a,LeeSai1987b}, alternative treatments have emerged
over the past decade \citep[see, \eg,][]{Lee1998,Lee2001}; however,
these are characterized by such a high degree of mathematical and
numerical complexity, that their \emph{large-scale} application is --
with present-day computational resources -- impractical.

The approach adopted herein aims at a compromise between
sophistication and practicality, by employing an approximate method to
treat the pulsation--rotation interaction (Sec.~\ref{sec:method}) that
is well-suited to the g modes found in SPB stars. This approach is
applied to a range of B-type stellar models (Sec.~\ref{sec:models}),
with the results of these stability calculations presented in
Section~\ref{sec:calc}. The findings of the analysis are then
discussed and summarized in Section~\ref{sec:summary}.


\section{Method} \label{sec:method}

\subsection{Theoretical Treatment} \label{ssec:theory}

From a theoretical standpoint, the introduction of rotation
significantly complicates the analysis of stellar pulsation. The
centrifugal force tends to distort the equilibrium star away from a
spherical configuration, and the Coriolis force leads to mixing
between the radial and angular components of the fluid momentum. Both
of these processes mean that the linearized equations describing
nonradial, nonadiabatic pulsation are no longer separable in all
coordinates, leading to a significant increase in the computational
effort necessary for their solution \citep[see, \eg,][]{Lee2001}.

However, for the case of the low-frequency g modes found in SPB stars,
an approximate treatment of the Coriolis force can be used to restore
the separability of the pulsation equations, and thereby render the
problem tractable even with modest computing resources. At the heart
of this approach is the so-called `traditional approximation',
introduced in a geophysical context by \citet{Eck1960}, and first
applied to stellar pulsation by \citet{LeeSai1987a}. The traditional
approximation involves neglecting the polar component of the rotation
angular frequency vector \rfreqv, in the Coriolis terms of the
linearized momentum equation. As discussed by \citet{LeeSai1997}, this
approximation is reasonable throughout regions where the pulsation
frequency \pfreq\ (as measured in the corotating frame) and the
rotation frequency \rfreq\ are both very much less than the \BV\
frequency \brunt\ associated with the gravitational stratification of
the medium. In the case of g modes in B-type stars, this condition is
fulfilled in almost all of the stably stratified, radiative envelope
where the modes are propagative.

In addition to the traditional approximation, the treatment presented
herein requires a number of other simplifications to permit the
separation of the pulsation equations in all three spherical-polar
coordinates $\{r,\theta,\phi\}$. These simplifications, which follow
the analysis by \citet{Tow2003b}, are:
\begin{enumerate}
\item The assumption that the rotation\footnote{Assumed throughout to
  be uniform.} is slow enough for both the centrifugal force, and the
  departures from sphericity engendered by it, to be neglected. In
  this context, `slow' may be interpreted as the condition $\rfreq^{2}
  \ll \rfreqc^{2}$, where
  \begin{equation}
    \rfreqc \equiv \sqrt{\frac{8 G\Mstar}{27 \Rstar^{3}}}
  \end{equation}
  defines the critical rotation frequency within the Roche model, with
  \Mstar\ and \Rstar\ being the stellar mass and radius, and $G$ the
  gravitational constant.
\item The \citet{Cow1941} approximation, whereby perturbations to the
  gravitational potential are neglected. This approximation is
  reasonable for all but low-order, low-degree modes.
\item The nonadiabatic radial flux (NARF) approximation, whereby the
  divergence of the horizontal Eulerian flux perturbation is neglected
  in the energy equation. As demonstrated by \citet{Tow2003b}, this
  approximation -- first introduced by \citet{Sav1995} -- is valid
  under the same conditions ($\pfreq,\rfreq \ll \brunt$) as the
  traditional approximation itself.
\item The steady-wave approximation, whereby the imaginary part
  \pfreqi\ of the pulsation frequency \pfreq\ is neglected in solving
  the angular parts of the pulsation equations. This approximation,
  which is discussed by \citet{Tow2000}, is appropriate when the
  growth rate of the pulsation is slow, so that \pfreqi\ is very much
  smaller than the corresponding real part \pfreqr\ of the pulsation
  frequency.
\end{enumerate}

This collection of approximations and assumptions closely follows
those employed by \citet{UshBil1998}, with the exception that their
use of the quasi-adiabatic approximation is replaced herein by the
adoption of the NARF approximation. This replacement is particularly
significant, in that the quasi-adiabatic approximation can lead to
incorrect results when applied to B-type stars \citep[see,
\eg,][]{Dzi1993}; in contrast, the NARF approximation is able to
capture the essential physics of the nonadiabatic processes
responsible for the $\kappa$-mechanism instability.

\subsection{Pulsation Equations} \label{ssec:equations}

Within the approximations outlined in the preceding section, the
spatial and temporal dependence of perturbed variables is expressed in
the corotating frame as
\begin{align}
f'(r,\mu,\phi,t) &= f'(r)\, \hough(\mu)\, \ee^{\ii(m \phi + \pfreq t)}\\
\intertext{and}
\delta f(r,\mu,\phi,t) &= \delta f(r)\, \hough(\mu)\, \ee^{\ii(m \phi + \pfreq t)},
\end{align}
where Eulerian and Lagrangian perturbations are represented by primes
($'$) and $\delta$, respectively; $t$ is the temporal coordinate; and
$\mu \equiv \cos\theta$. Here, the Hough function $\hough(\mu)$ is a
solution to Laplace's tidal equation \citep[see, \eg,][]{Bil1996},
that depends implicitly on the `spin parameter' $\nu \equiv
2\rfreq/\pfreqr$ characterizing the relative strength of Coriolis and
buoyancy forces \citep{Tow2005}. In the limit of no rotation,
$\hough(\mu)$ approaches the normalized Legendre function $\nPml(\mu)$
\citep[see][]{LeeSai1990} having harmonic degree $\ell$ and azimuthal
order $m$.

Inspired by the dimensionless formulations introduced by
\citet{SaiCox1980} and \citet{LeeSai1997}, which themselves are built
on the seminal \citet{Dzi1971} treatment, the radial dependencies
$f'(r)$ and $\delta f(r)$ of perturbed variables are expressed in
terms of a set of eigenfunctions $q_{i}$ ($i=1,2,5,6$)\footnote{The
indices 3 and 4 are traditionally reserved for the perturbation to the
gravitational potential and its radial derivative; due to the adoption
of the \citet{Cow1941} approximation, they are not used in the present
analysis.}, such that
\begin{align} \label{eqn:eigen}
\frac{\xir(r)}{r} &= q_{1}\,x^{\elle - 2},      &
\frac{p'(r)}{\rho g r} &= q_{2}\,x^{\elle - 2}, \\
\frac{\delta S(r)}{c_{p}} &= q_{5}\,x^{\elle},  &
\frac{\delta \Lrad(r)}{\Lstar} &= q_{6}\,x^{\elle + 1}.
\end{align}
Here, $x \equiv r/\Rstar$ is the dimensionless radial coordinate,
\xir\ the radial fluid displacement, $p'$ the Eulerian pressure
perturbation, $\delta S$ the Lagrangian perturbation to the specific
entropy, and $\delta \Lrad$ the Lagrangian perturbation to the radial
part of the radiative luminosity. Other symbols appearing in these
expressions are defined by \citet{Unn1989}, with the exception of the
effective harmonic degree
\begin{equation} \label{eqn:elle}
\elle = \frac{\sqrt{1 + 4\lamml} - 1}{2},
\end{equation}
introduced by \citet{Tow2000} as the solution of the equation
$\elle(\elle+1) = \lamml$. In these relations, \lamml\ is the
eigenvalue of Laplace's tidal equation associated with the Hough
eigenfunction $\hough(\mu)$. This eigenvalue depends implicitly on the
spin parameter $\nu$, and therefore on both the pulsation frequency
\pfreq\ and the rotation frequency \rfreq.

The radial eigenfunctions $q_{i}$ are found as the solutions of four
coupled, first-order differential equations,
\begin{gather} \label{eqn:puls-disp}
x \frac{\diff q_{1}}{\diff x} =
\left(\frac{V}{\Gamma_{1}} - 1 - \elle \right) q_{1} +
\left[\frac{\elle(\elle+1)}{c_{1}\dfreq^{2}} - \frac{V}{\Gamma_{1}}\right] q_{2} +
\upsilon_{T} x^{2} q_{5}, \\ \label{eqn:puls-pres}
x \frac{\diff q_{2}}{\diff x} =
(c_{1}\dfreq^{2} - A^{\ast}) q_{1} +
\left(A^{\ast} - U + 3 - \elle \right) q_{2} +
\upsilon_{T} x^{2} q_{5}, \\ \label{eqn:puls-radd}
\begin{split}
x \frac{\diff q_{5}}{\diff x} &= 
  V \left[ \nablad(U - c_{1}\dfreq^{2}) -  4(\nablad - \nabla) + c_{2}
  \right] \frac{q_{1}}{x^{2}} + \mbox{} \\
& V \left[ \frac{\elle(\elle+1)}{c_{1}\dfreq^{2}}(\nablad - \nabla) - c_{2}
  \right] \frac{q_{2}}{x^{2}} + \mbox{} \\
& \left[ V \nabla \left(4 - \kappa_{S}\right) - \elle \right] q_{5} - \frac{V
  \nabla}{\lrad} x q_{6},
\end{split} \\ \label{eqn:puls-enrg}
\begin{split}
x \frac{\diff q_{6}}{\diff x} &=
  - \epsad \ceps V \frac{q_{1}}{x^{3}} + \left[ \epsad \ceps V +
    \frac{\elle(\elle+1)}{c_{1}\dfreq^{2}} x \frac{\diff \lrad}{\diff x}
    \right] \frac{q_{2}}{x^{3}} + \mbox{} \\
& \left( \ceps \epss - \ii \dfreq \cthr \right) \frac{q_{5}}{x} - \left( \elle +
    1 \right) q_{6}.
\end{split}
\end{gather}
Here, the dimensionless pulsation frequency \dfreq\ has the usual
definition
\begin{equation}
\omega \equiv \sigma\,\sqrt{\frac{\Rstar^{3}}{G \Mstar}},
\end{equation}
and all other symbols follow the nomenclature of \citet{Unn1989}, with
the exception of the introductions
\begin{gather}
\lrad \equiv \frac{\Lrad}{\Lstar}, \\
\ceps \equiv c_{3}\,\lrad = \frac{4\pi r^{3} \rho \epsn}{\Lstar}, \\
\intertext{and}
\cthr \equiv c_{4}\,\lrad = \frac{4\pi r^{3} \rho T c_{p}}{\Lstar} \sqrt{\frac{G \Mstar}{\Rstar^{3}}}.
\end{gather}

In the limit $\nu \rightarrow 0$, the effective harmonic degree \elle\
approaches the true harmonic degree $\ell$, and the governing
equations~(\ref{eqn:puls-disp}--\ref{eqn:puls-enrg}) reduce to those
describing nonradial, nonadiabatic pulsation of a non-rotating star,
within the NARF and \citet{Cow1941} approximations. However, even
allowing for the alternative nomenclature, the energy conservation
equation~(\ref{eqn:puls-enrg}) in this limit appears rather different
than in other nonadiabatic treatments \citep[compare with, \eg,][their
eqn.~24.12\footnote{Note that the sign of the $c_{3}/c_{1}\dfreq^{2}$
term in their equation is incorrect.}]{Unn1989}. In particular, the
derivative of the dimensionless radiative luminosity \lrad\ appears in
the present treatment due to the choice of `frozen convection'
approximation; this choice centres around neglecting perturbations to
the convective source term in the linearized energy equation, as per
eqn.~21.7 of \citet{Unn1989}. Although \citet{Pes1990} has argued that
such an approach is unphysical, it is required to ensure the pulsation
equations remain self-consistent at the origin. In any case, since g
modes do not penetrate far into convective regions, any errors
introduced by this convection-freezing choice should be minimal.

Solutions of the pulsation
equations~(\ref{eqn:puls-disp}--\ref{eqn:puls-enrg}) are required to
satisfy boundary conditions at the centre and at the surface. The
inner conditions
\begin{equation} \label{eqn:bound-inner}
\left.
\begin{aligned}
c_{1}\dfreq^{2} \,q_{1} - \elle \,q_{2} &= 0 \\
3 \lrad \, q_{1} - x^{3} q_{6} &= 0
\end{aligned}
\right\} \quad \text{as $x \rightarrow 0$},
\end{equation}
ensure that solutions remain finite and regular at the
origin. Likewise, the outer boundary conditions
\begin{equation} \label{eqn:bound-outer}
\left.
\begin{aligned}
q_{1} - q_{2} &= 0 \\
(2 - 4\nablad V)\,q_{1} + 4\nablad V\,q_{2} + 4\,q_{5} - q_{6} &= 0
\end{aligned}
\right\} \quad \text{at $x = 1$}
\end{equation}
follow from the assumptions that the surface pressure tends to zero,
and that there is no flux incident from outside the star \citep[see,
\eg,][]{Unn1989}. Finally, the arbitrary overall scaling of solutions
is constrained by the normalization condition
\begin{equation} \label{eqn:bound-norm}
q_{1} = 1 \quad \text{at $x=1$}.
\end{equation}

\subsection{Implementation} \label{ssec:implement}

To solve the pulsation
equations~(\ref{eqn:puls-disp}--\ref{eqn:puls-enrg}) and accompanying
boundary conditions~(\ref{eqn:bound-inner}--\ref{eqn:bound-norm}), a
wholly-new Fortran 95 code was developed. The code, named \boojum,
follows the root-finding approach pioneered by \citet{Cas1971} and
\citet{OsaHan1973}: one of the boundary conditions is set aside,
allowing solution of the equations to be achieved at arbitrary
dimensionless frequency \dfreq. The suppressed boundary condition is
then used to construct a discriminant $\disc(\dfreq)$, whose roots
correspond to the eigenfrequencies of the full system of equations.

In the present case, the outer mechanical boundary condition is used
to form the discriminant
\begin{equation}
\disc(\dfreq) = \frac{\left(q_{1} - q_{2}\right)_{x=1}}{\left(q_{1} +
  q_{2}\right)_{x=0}}.
\end{equation}
The numerator is the boundary condition itself, while the denominator
-- which is guaranteed by the inner boundary
conditions~(\ref{eqn:bound-inner}) never to be zero -- ensures that
the discriminant remains well behaved \citep[see][]{Tow2000}. To
calculate the solution vector $q_{i}$ required to evaluate this
discriminant, \boojum\ uses a standard relaxation approach
\citep[\eg,][]{Pre1992}; because the pulsation equations are linear in
$q_{i}$, only a single iteration is required at each value of \dfreq.
The \elle-dependent scalings, adopted in the
definitions~(\ref{eqn:eigen}) of the eigenfunctions, successfully
prevent the loss of precision near the centre discussed by
\citet{TakLof2004}. For improved accuracy, centred finite differences
are used in the relaxation algorithm; however, in regions where $\cthr
> 10^{4}$, \boojum\ switches to one-sided differences for the thermal
equations~(\ref{eqn:puls-radd}--\ref{eqn:puls-enrg}), in accordance
with the \citet{Sug1970} prescription for avoiding numerical stability
\citep[and see also][their \S24]{Unn1989}.

Solution of the characteristic equation
\begin{equation} \label{eqn:char}
\disc(\dfreq) = 0,
\end{equation}
defining the eigenfrequencies of the pulsation equations, is
accomplished in \boojum\ using the robust \citet{Tra1964}
implementation of the root-finding algorithm devised by
\citet{Mul1956}; this algorithm is a generalization of the complex
secant approach favoured by \citet{Cas1971}. Note that the latter
author used approximate roots of the characteristic equation to find
starting points for the solution, via iterative relaxation, of the
full pulsation equations. A somewhat simpler approach is adopted by
\boojum, whereby equation~(\ref{eqn:char}) is solved to the desired
fractional tolerance in \dfreq\ (in the present work, $10^{-10}$ for
both real and imaginary parts), and no additional calculations are
performed.

For calculation of the \lamml\ eigenvalue, required (\cf\
eqn.~\ref{eqn:elle}) to evaluate the \elle\ terms appearing in the
pulsation equations and boundary conditions, \boojum\ uses the
matrix-mechanical approach pioneered by \citet{LeeSai1987a}. The
implementation largely follows the procedure described by
\citet{Tow2003b}; however, the truncation dimension $N$ is determined
dynamically, by repeatedly doubling $N$ until the fractional change in
\lamml\ drops below some specified threshold (taken to be $10^{-5}$
throughout the present work). Furthermore, the algorithm due to
\citet{Kah1966}, as implemented by the \textsc{lapack} subroutine
library \citep{And1999}, is used for calculating the matrix
eigenvalues; this algorithm performs significantly better at finding
isolated eigenvalues than the previously-adopted \emph{QL} approach.

As input, \boojum\ is supplied with the desired mode parameters
$(\ell,m)$, the rotation angular frequency \rfreq, and an indication
of the region over which to search for eigenfrequencies satisfying the
characteristic equation~(\ref{eqn:char}). The coefficients appearing
in the pulsation equations and boundary conditions are evaluated using
a precomputed stellar model (see Sec.~\ref{sec:models}); the model --
typically composed of \ca\ 1,000 radial points -- is interpolated onto
a new grid having 10,000 points, distributed non-uniformly via an
approach similar to that described by \citet[][their Appendix
A3]{ChrMul1994}. Cubic splines are used for interpolating all
variables apart from the \BV\ frequency \brunt; the latter is
interpolated linearly, to avoid the introduction of spurious
oscillations in the molecular weight gradient zone situated outside
the convective core.

\section{Stellar Models} \label{sec:models}

\begin{figure}
\epsffile{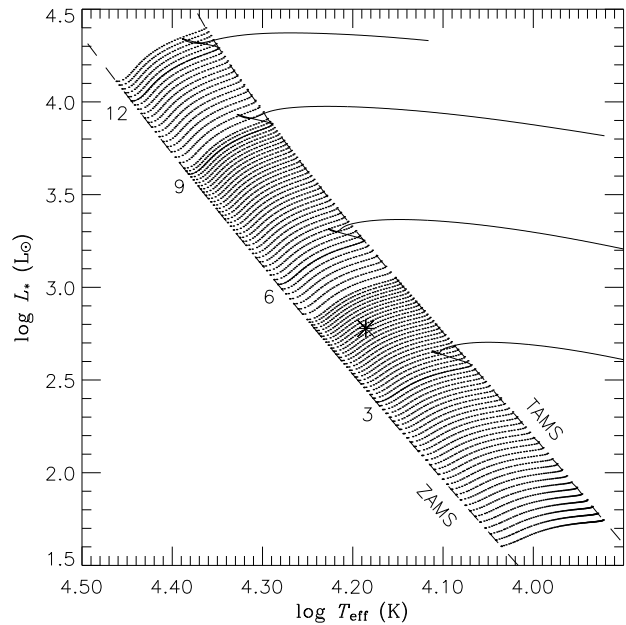}
\caption{The theoretical HR diagram for the stellar models introduced
in Section~\ref{sec:models}; these are plotted as points in the
effective temperature (\Teff) versus stellar luminosity (\Lstar)
plane. The dashed lines running diagonally from top left to bottom
right indicate the ZAMS and TAMS main sequence limits, while the
asterisk shows the location of the 53 Per-like model analyzed in
Section~\ref{ssec:53per}. Full evolutionary tracks for four selected
models are also shown in the diagram, plotted as solid lines and
labeled by their corresponding stellar masses, in solar units. The two
higher-mass tracks do not reach the low-temperature boundary of the
diagram, because the Warsaw-New Jersey evolutionary code is unable to
follow these models beyond the ignition of core helium burning.}
\label{fig:models}
\end{figure}

The Warsaw-New Jersey stellar evolution code is used to calculate 115
tracks of stellar models, sampling the initial mass range $\Mstar =
2.5\,\Msun$--$5.2\,\Msun$ at a resolution $0.05\,\Msun$, the range
$\Mstar = 5.2\,\Msun$--$9.2\,\Msun$ at a resolution $0.1\,\Msun$, and
the range $\Mstar = 9.2$--$13.0\,\Msun$ at a resolution $0.2\,\Msun$;
each track extends from zero-age main sequence (ZAMS) to terminal-age
main sequence (TAMS). Details of the code have already been given by
\citet{DziPam1993} and \citet{Dzi1993}; the only significant
difference in the present work is the adoption of more-recent
\textsc{opal} tabulations for opacity \citep{IglRog1996} and equation
of state \citep{Rog1996}. In all cases, the initial hydrogen and metal
mass fractions are set at the canonical values $X=0.7$ and $Z=0.02$,
with a heavy-element mixture taken from \citet{GreNoe1993}. No account
is taken of the effect of rotation on the stellar evolution; although
\citet{Pam1999} has demonstrated that rotation-induced mixing can
shift the red edge of the SPB instability strip to lower effective
temperatures, the present analysis is more concerned with the
\emph{dynamical} influence of rotation on SPB pulsation.

Each of the calculated evolutionary tracks is composed of
approximately 42 stellar models, with a grand total of 4,876 models
for the entire set. Fig.~\ref{fig:models} plots the positions of these
models in the theoretical Hertzsprung-Russell (HR) diagram, along with
the loci defining the ZAMS and TAMS boundaries. The dense model
coverage of the main sequence band is to facilitate the accurate
positioning of instability strips (Sec.~\ref{ssec:instab}) in the
$\Teff-\Lstar$ plane.

\section{Stability Calculations} \label{sec:calc}

\subsection{53 Per Model} \label{ssec:53per}

\begin{table}
\leavevmode
\begin{center}
\caption{Fundamental parameters of the 53 Per-like stellar model
analyzed in Section~\ref{ssec:53per}.} \label{tab:53per}
\begin{tabular}{@{}ccccc}
$\Teff/\Kelv$ & $\log g/\cmss$ & $\Lstar/\Lsun$ & $\Mstar/\Msun$ & $\Rstar/\Rsun$ \\ \hline
15,300        & 4.04           & 600            & 4.80           & 3.48
\end{tabular}
\end{center}
\end{table}

\begin{figure*}
\leavevmode
\begin{center}
\epsffile{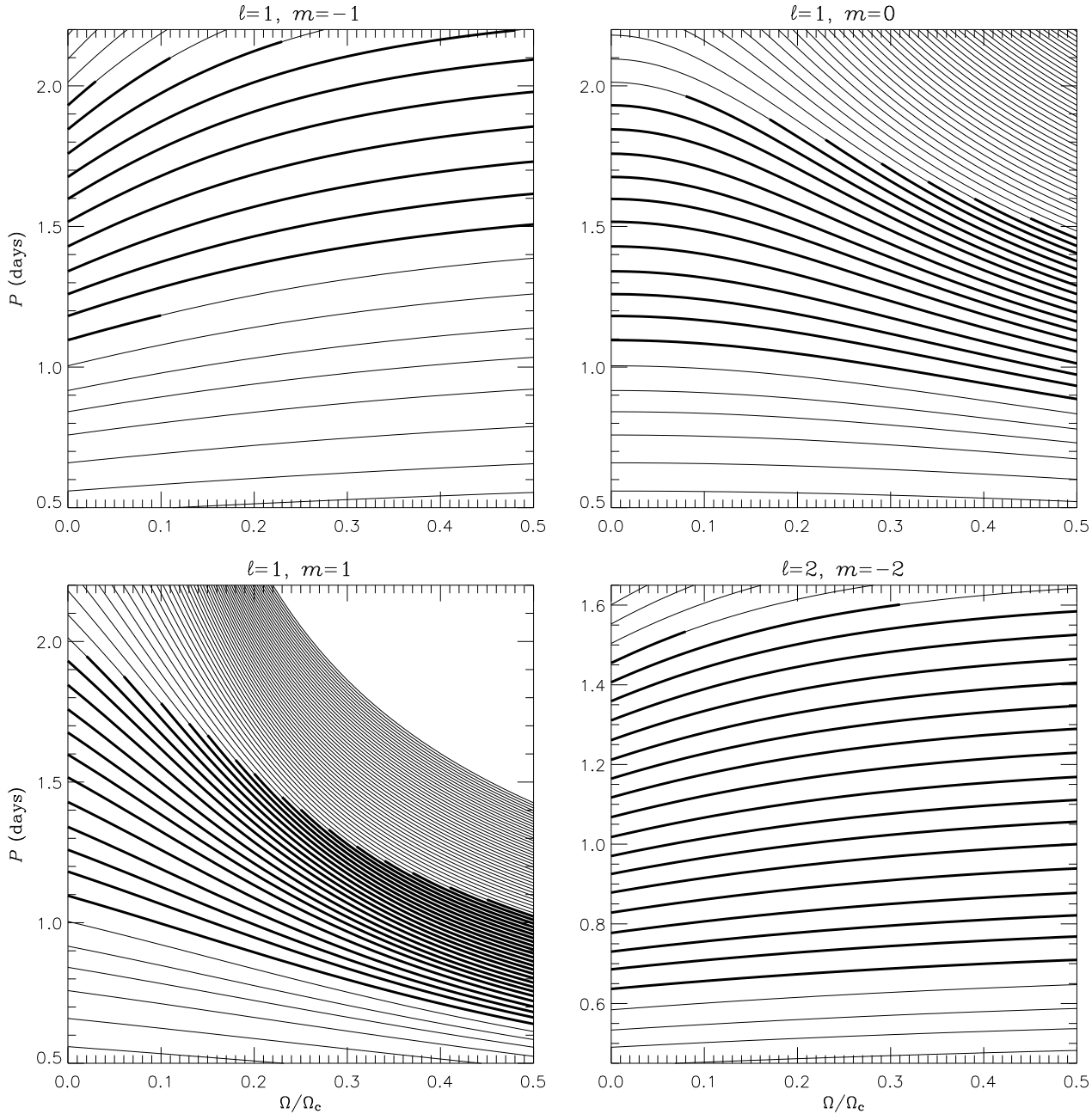}
\caption{Pulsation periods for $\ell=1$ and $\ell=2$ g modes of the 53
Per-like model, plotted as a function of rotation frequency up to a
maximum $\rfreq = 0.5\,\rfreqc$. Each curve corresponds to a
particular radial order \ncowl; the line weight is used to indicate
whether modes are stable (thin) or unstable (thick) against the
$\kappa$-mechanism instability. To improve the clarity of the figure,
only radial orders $|\ncowl| \leq 75$ are plotted.}
\label{fig:53per}
\end{center}
\end{figure*}

\begin{figure*}
\leavevmode
\begin{center}
\epsffile{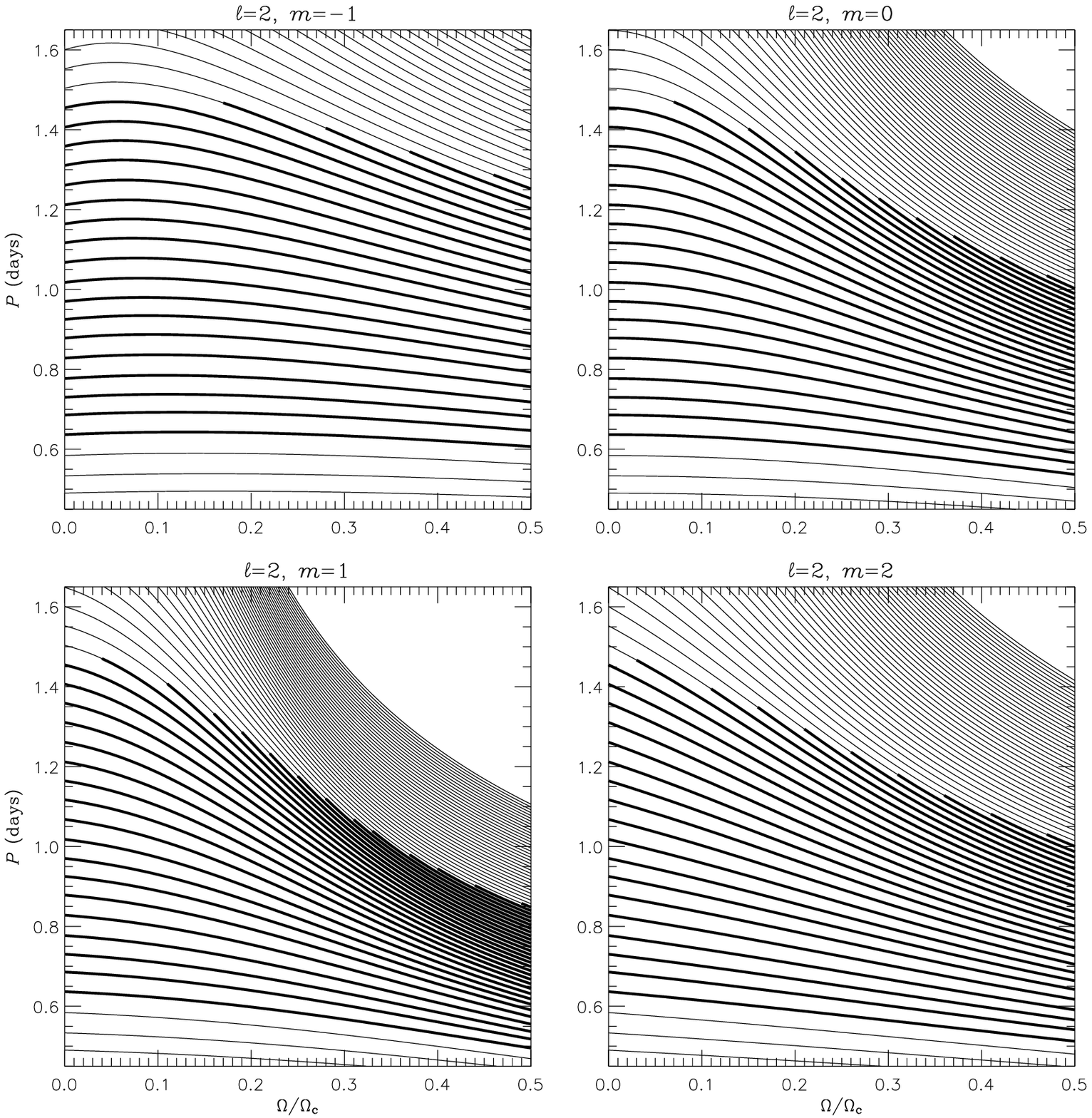}
\contcaption{}
\end{center}
\end{figure*}

From the models introduced in the preceding section, one is selected
as having parameters close to the values inferred by \citet{DeR1999}
for 53 Per, the archetypal SPB star; these parameters are documented
in Table~\ref{tab:53per}, and the corresponding position in the HR
diagram is shown in Fig.~\ref{fig:models} by an asterisk. This 53
Per-like model is employed to examine the general influence the
Coriolis force exerts over $\ell=1$ and $\ell=2$ nonradial g
modes. Eigenfrequencies of these modes are calculated using \boojum,
for all possible values $-\ell \leq m \leq \ell$ of the azimuthal
order, and over a range of angular frequencies ranging from the
non-rotating case up to the intermediate rate $\rfreq/\rfreqc =
0.5$. This upper limit corresponds to a rotation period $\rper =
1.26\,\days$, with an equatorial velocity \veq\ that is 35\% of the
star's critical velocity $\vcrit = 419\,\kms$; by way of comparison,
the most rapidly-rotating SPB star known to date, HD~1976, rotates at
$\sim 32\%$ of its critical velocity \citep{Mat2001}.

The results from these calculations are shown in Fig.~\ref{fig:53per},
where the corotating frame period $\pper \equiv 2\pi/\pfreqr$ of each
mode is plotted as a function of rotation angular frequency. Points
associated with eigenfunctions $q_{i}$ having the same radial order
\ncowl\footnote{As defined by \citet[][their eqn.~17.5]{Unn1989},
within the generalization to the \citet{Cow1941} nomenclature
introduced by \citet{Scu1974} and \citet{Osa1975}.} are linked
together into a single, continuous curve; the weight of the curves, at
each value of \rfreq, is used to indicate whether a mode is stable
($\pfreqi > 0$) or unstable ($\pfreqi < 0$) against $\kappa$-mechanism
excitation.

At $\rfreq=0$, the eigenfrequencies are degenerate in azimuthal order
$m$, owing to the arbitrariness of the model's polar axis. This
degeneracy is lifted upon the introduction of rotation, with a
splitting initially following the first-order relation
\begin{gather}
\Delta \pfreqr \approx m \rfreq \Cnl, \\ 
\intertext{or alternatively} 
\Delta \pper \approx - m \rper \Cnl,
\end{gather}
derived by \citet{Led1951}; here, $\rper\equiv2\pi/\rfreq$ is the
rotation period, and $\Delta$ denotes the change in the indicated
quantity. In the case of the high-order g modes considered herein, the
term \Cnl\ approximates to $1/\ell(\ell+1)$, explaining why the
$\ell=1$ modes exhibit steeper gradients $\Delta\pper/\Delta\rfreq$
around $\rfreq \approx 0$.

\begin{table}
\leavevmode
\begin{center}
\caption{Instability ranges for g modes of the 53 Per-like model; for
each $(\ell,m)$ pair considered, and at three differing rotation
rates, the range of $|\ncowl|$ spanned by unstable modes is tabulated,
along with the corresponding values of the corotating frame period
\pper, the inertial frame period \pperi\ and the harmonic degree ratio
$\elle/\ell$. The absolute value of \ncowl\ is used, because this
radial order is negative for g modes.}
\label{tab:modes}
\begin{tabular}{@{}ccccc}
$(\ell,m)$ &
$|\ncowl|$ & \pper\ (\days) & \pperi\ (\days) & $\elle/\ell$ \\ \hline
\\
\multicolumn{5}{c}{$\rfreq/\rfreqc = 0.00$} \\
$(1,-1)$ & 12 -- 22 & 1.10 -- 1.93 & 1.10 -- 1.93 & 1.00 -- 1.00 \\
$(1,0)$ & 12 -- 22 & 1.10 -- 1.93 & 1.10 -- 1.93 & 1.00 -- 1.00 \\
$(1,1)$ & 12 -- 22 & 1.10 -- 1.93 & 1.10 -- 1.93 & 1.00 -- 1.00 \\
$(2,-2)$ & 12 -- 29 & 0.64 -- 1.46 & 0.64 -- 1.46 & 2.00 -- 2.00 \\
$(2,-1)$ & 12 -- 29 & 0.64 -- 1.46 & 0.64 -- 1.46 & 2.00 -- 2.00 \\
$(2,0)$ & 12 -- 29 & 0.64 -- 1.46 & 0.64 -- 1.46 & 2.00 -- 2.00 \\
$(2,1)$ & 12 -- 29 & 0.64 -- 1.46 & 0.64 -- 1.46 & 2.00 -- 2.00 \\
$(2,2)$ & 12 -- 29 & 0.64 -- 1.46 & 0.64 -- 1.46 & 2.00 -- 2.00 \\
\\
\multicolumn{5}{c}{$\rfreq/\rfreqc = 0.25$} \\
$(1,-1)$ & 13 -- 19 & 1.40 -- 2.07 & 0.90 -- 1.14 & 0.79 -- 0.75 \\
$(1,0)$ & 12 -- 25 & 1.03 -- 1.77 & 1.03 -- 1.77 & 1.09 -- 1.32 \\
$(1,1)$ & 12 -- 31 & 0.85 -- 1.40 & 1.27 -- 3.16 & 1.40 -- 2.26 \\
$(2,-2)$ & 12 -- 28 & 0.68 -- 1.58 & 0.44 -- 0.70 & 1.84 -- 1.74 \\
$(2,-1)$ & 12 -- 30 & 0.64 -- 1.40 & 0.51 -- 0.90 & 1.99 -- 2.18 \\
$(2,0)$ & 12 -- 33 & 0.61 -- 1.28 & 0.61 -- 1.28 & 2.13 -- 2.70 \\
$(2,1)$ & 12 -- 35 & 0.58 -- 1.18 & 0.76 -- 2.21 & 2.22 -- 3.18 \\
$(2,2)$ & 12 -- 33 & 0.58 -- 1.23 & 1.07 -- 46.50 & 2.25 -- 2.83 \\
\\
\multicolumn{5}{c}{$\rfreq/\rfreqc = 0.50$} \\
$(1,-1)$ & 13 -- 18 & 1.51 -- 2.10 & 0.69 -- 0.79 & 0.72 -- 0.69 \\
$(1,0)$ & 12 -- 29 & 0.89 -- 1.46 & 0.89 -- 1.46 & 1.32 -- 2.00 \\
$(1,1)$ & 12 -- 38 & 0.64 -- 1.04 & 1.30 -- 5.76 & 1.99 -- 4.02 \\
$(2,-2)$ & 12 -- 27 & 0.71 -- 1.59 & 0.33 -- 0.45 & 1.75 -- 1.66 \\
$(2,-1)$ & 12 -- 33 & 0.61 -- 1.25 & 0.41 -- 0.63 & 2.12 -- 2.77 \\
$(2,0)$ & 12 -- 38 & 0.54 -- 1.00 & 0.54 -- 1.00 & 2.46 -- 4.17 \\
$(2,1)$ & 12 -- 42 & 0.50 -- 0.85 & 0.82 -- 2.59 & 2.71 -- 5.58 \\
$(2,2)$ & 12 -- 38 & 0.51 -- 1.00 & 2.73 -- -1.71 & 2.60 -- 4.18 \\
\end{tabular}
\end{center}
\end{table}

Toward larger values of \rfreq, departures from the above linear
relations are increasingly apparent in Fig.~\ref{fig:53per}. For all
apart from the prograde sectoral (PS) modes having $\ell=-m$, these
departures produce an overall trend of decreasing pulsation period
with increasing rotation rate, irrespective of whether the mode is
prograde ($m<0$), retrograde ($m>0$) or zonal ($m=0$). Such behaviour
comes from the influence of the Coriolis force, and can readily be
understood by recalling that -- at the most general level -- the
frequency of a wave may be expressed as the square root of the ratio
between a generalized stiffness and a generalized inertia \citep[see,
\eg,][]{Lig1978}. With the introduction of rotation, the
stiffness\footnote{That is, the restoring force on displaced fluid
elements.} usually due to buoyancy is augmented by the
Coriolis force; this leads to an increase in frequency, and thence a
corresponding \emph{decrease} in mode period.

Because the strength of the Coriolis force varies with spin parameter
$\nu$, the effects described are differential: the periods of
long-period, larger-$\nu$ modes are shortened by a far greater degree
than those of short-period, smaller-$\nu$ modes. Accordingly, the
density of the pulsation spectrum, as measured by the number of modes
either per period interval or per frequency interval, increases
markedly with rotation rate. As can be seen in Fig.~\ref{fig:53per},
this \emph{spectral compression} is especially pronounced for the
$m=1$ modes.

This latter result warrants some explanation. For non-PS modes in the
inertial regime $|\nu| > 1$, the Coriolis force acts to confine the
modes within an equatorial waveguide, whose boundaries are situated at
$\mu = \pm 1/|\nu|$ \citep[see, \eg,][]{Bil1996}. A requirement of the
trapping is that the $s+1$ nodes of the Hough function \hough\ be
fitted within these waveguide boundaries, where
\begin{equation}
s = \ell - |m| \pm 1
\end{equation}
is the meridional order introduced by \citet{Tow2003a}. In this
expression, the difference between the prograde and zonal modes ($m
\leq 0$; minus sign) and the retrograde modes ($m >0 $; plus sign)
comes about because an extra pair of nodes appears in the latter when
$|\nu|>1$ \citep[see][]{LeeSai1990}. In the limit $|\nu|\gg1$, the
fitting requirement is embodied in the angular characteristic equation
\begin{equation}
\lamml \approx \nu^{2}(2s + 1)^{2}
\end{equation}
for the Hough-function eigenvalue \citep[][his eqn.~38]{Tow2003a},
corresponding to an effective harmonic degree
\begin{equation} \label{eqn:elle-asymp}
\elle \approx \sqrt{\lamml} \approx 
\left\{
\begin{aligned}
(2\ell - 2|m| - 1)\,\nu & \qquad (m \leq 0), \\
(2\ell - 2|m| + 3)\,\nu & \qquad (m > 0).
\end{aligned}
\right.
\end{equation}
Regarding the ratio $\elle/\ell$ as a measure of the severity of
equatorial confinement, this latter expression indicates $m=1$ modes
-- for each value of $\ell$ -- are the most affected by the Coriolis
force. This result can be seen in the right-most column of
Table~\ref{tab:modes} introduced below; physically, it follows
directly from the fact that these modes have the largest meridional
orders ($s=1$ for $\ell=1$, and $s=2$ for $\ell=2$,) and are therefore
the most compressed in the polar direction when squeezed into the
equatorial waveguide.

The foregoing analysis does not apply to the PS modes,
$(\ell,m)=(1,-1)$ and $(\ell,m)=(2,-2)$, whose periods \emph{lengthen}
somewhat toward larger \rfreq. As discussed by \citet{Tow2003a}, this
class of mode is transformed by the Coriolis force into equatorial
Kelvin waves. Such waves have different properties than the usual
gravito-inertial waves found in a rotating, stratified system; in
particular, they are characterized by geostrophic balance, whereby the
Coriolis force arising from azimuthal fluid motions is countered by
polar pressure gradients \citep[see, \eg][]{Gil1982}. Due to this
balance, the Coriolis force makes relatively little difference to the
generalized stiffness of PS modes, and therefore does not produce the
marked period decrease seen for the other modes.

As a prelude to the analysis presented in the following section, the
focus now turns briefly toward the \emph{stability} of the modes
plotted in Fig.~\ref{fig:53per}. Once again, there is a dichotomy
between the PS modes and the others: the former are partly stabilized
by rotation, and the latter partly destabilized. However, in both
cases the rotation does not alter the property that a
\emph{contiguous} sequence of modes is unstable toward the $\kappa$
mechanism. With this result in mind, Table~\ref{tab:modes} succinctly
summarizes the data shown in the figure, by indicating the range of
the instability, at three differing rotation rates, for each
$(\ell,m)$ pair considered. This range is expressed in terms of the
radial orders of the highest- and lowest-frequency unstable modes; the
corresponding corotating frame period \pper, inertial frame
period\footnote{For retrograde modes, negative values of \pperi\ --
\eg, in the final row of Table~\ref{tab:modes} -- indicate the mode is
\emph{prograde} in the inertial frame.} $\pperi \equiv 2\pi/(\pfreq -
m \rfreq)$, and harmonic degree ratio $\elle/\ell$ are also
tabulated. From the table, it is evident that a grand total of 7 out
of the 29 initially unstable PS modes are stabilized by the
rotation. In the same way, the 94 non-PS modes that are initially
unstable are augmented, upon the introduction of rotation, by the
destabilization of a further 58; the largest gains, of 16 additional
unstable modes, are accorded to the $(\ell,m)=(1,1)$ group.

An important result clearly seen in Fig.~\ref{fig:53per}, is that the
stabilization or destabilization of modes, with varying rotation rate,
does not suffice to maintain the boundaries of the instability at a
constant frequency or period. For instance, the $(\ell,m)=(1,1)$ modes
mentioned above exhibit instability over the period range $\sim
1.05$--$1.95\,\days$ in the non-rotating case; but this range is
narrowed and shifted to $\sim 0.65$--$1.05\,\days$ in the
rapid-rotation limit. Thus, even though \emph{more} modes are unstable
in this limit, they occupy a narrower region of the pulsation
spectrum; such behaviour is a direct consequence of the spectral
compression discussed previously.

The processes selecting which modes are unstable are similar to those
operative in a non-rotating star. As discussed by \citet{Dzi1993}, one
requirement for the $\kappa$ mechanism to work efficiently is that the
relative Lagrangian pressure perturbation $\delta p/p$ be large and
slowly varying with radius within the excitation zone. This condition
establishes the short-period limit of the instability at $|\ncowl|
\sim 12$ (\cf\ Table~\ref{tab:modes}); because the shape of radial
eigenfunctions is insensitive to the strength of the Coriolis force
\citep[see, \eg,][]{UshBil1998}, this limit does not vary appreciably
with rotation rate.

The corresponding long-period limit is determined not so much by the
efficiency of the $\kappa$-mechanism driving, but more by the onset of
significant thermal damping. This damping arises due to radiative
diffusion between neighbouring fluid elements, operating primarily in
a region of the envelope ($\log T \sim 5.9$) situated below the
excitation zone. For the diffusion to be effective in stabilizing the
pulsation requires a combination of long pulsation periods and high
radial orders, the latter serving to steepen the temperature gradients
driving the diffusion. The interplay between these two factors
leads to a long-period instability limit that is quite sensitive to
the rotation rate, as Fig.~\ref{fig:53per} attests.

\subsection{Instability Strips} \label{ssec:instab}

\begin{figure*}
\leavevmode
\begin{center}
\epsffile{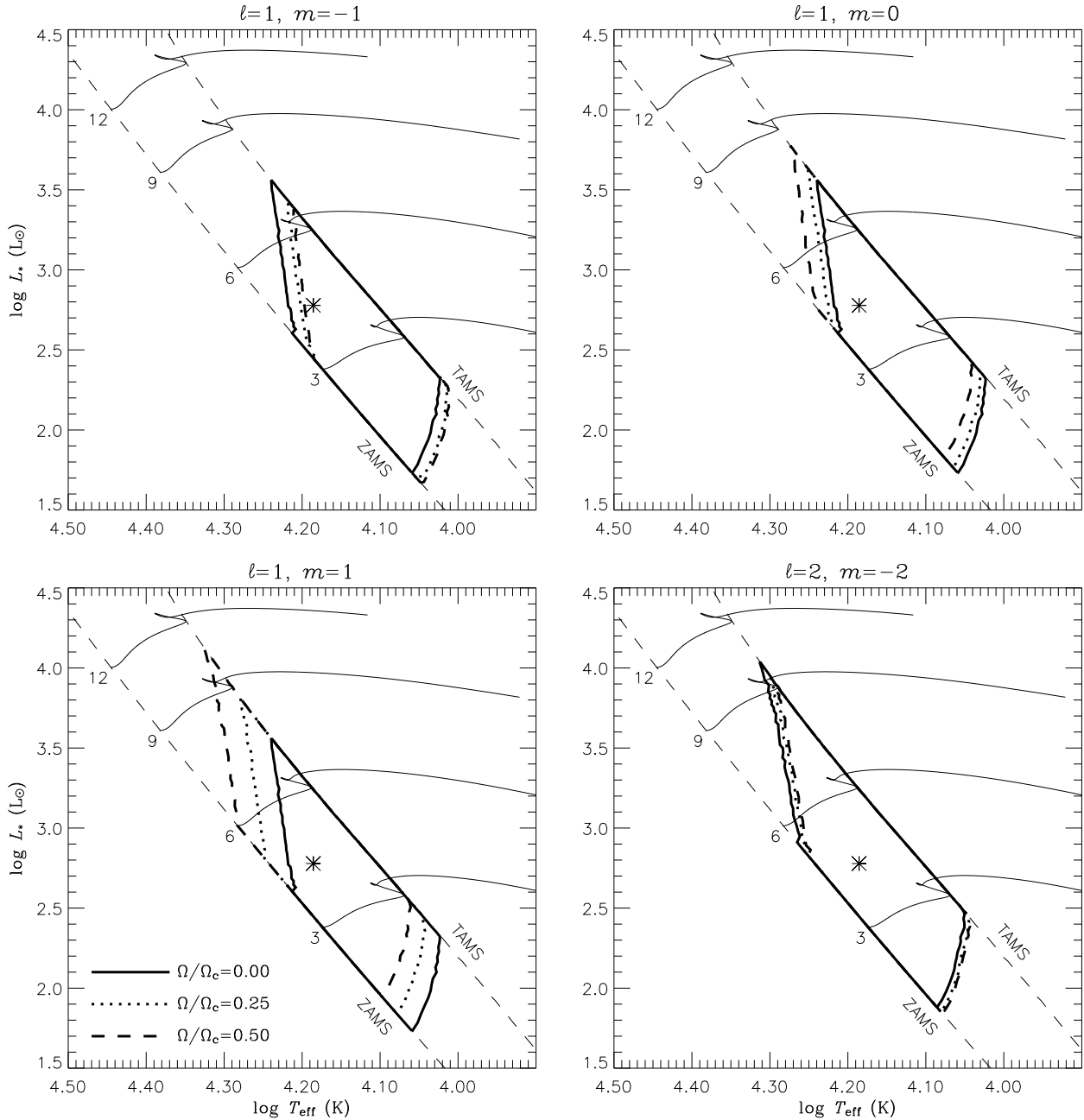}
\caption{Instability strips in the HR diagram, for $\ell=1$ and
$\ell=2$ g modes of the stellar models introduced in
Fig.~\ref{fig:models}. In each panel, corresponding to a particular
combination $(\ell,m)$ of mode parameters, the extent of the
$\kappa$-mechanism instability is indicated using thick lines, at the
three differing rotation rates $\rfreq/\rfreqc=0.0$ (solid), 0.25
(dotted) and 0.5 (dashed) considered. The uneven boundaries of the
instability strips, evident in some panels, are due to the discrete
spacing of the stellar models in the \Teff--\Lstar\ plane.}
\label{fig:instab}
\end{center}
\end{figure*}

\begin{figure*}
\leavevmode
\begin{center}
\epsffile{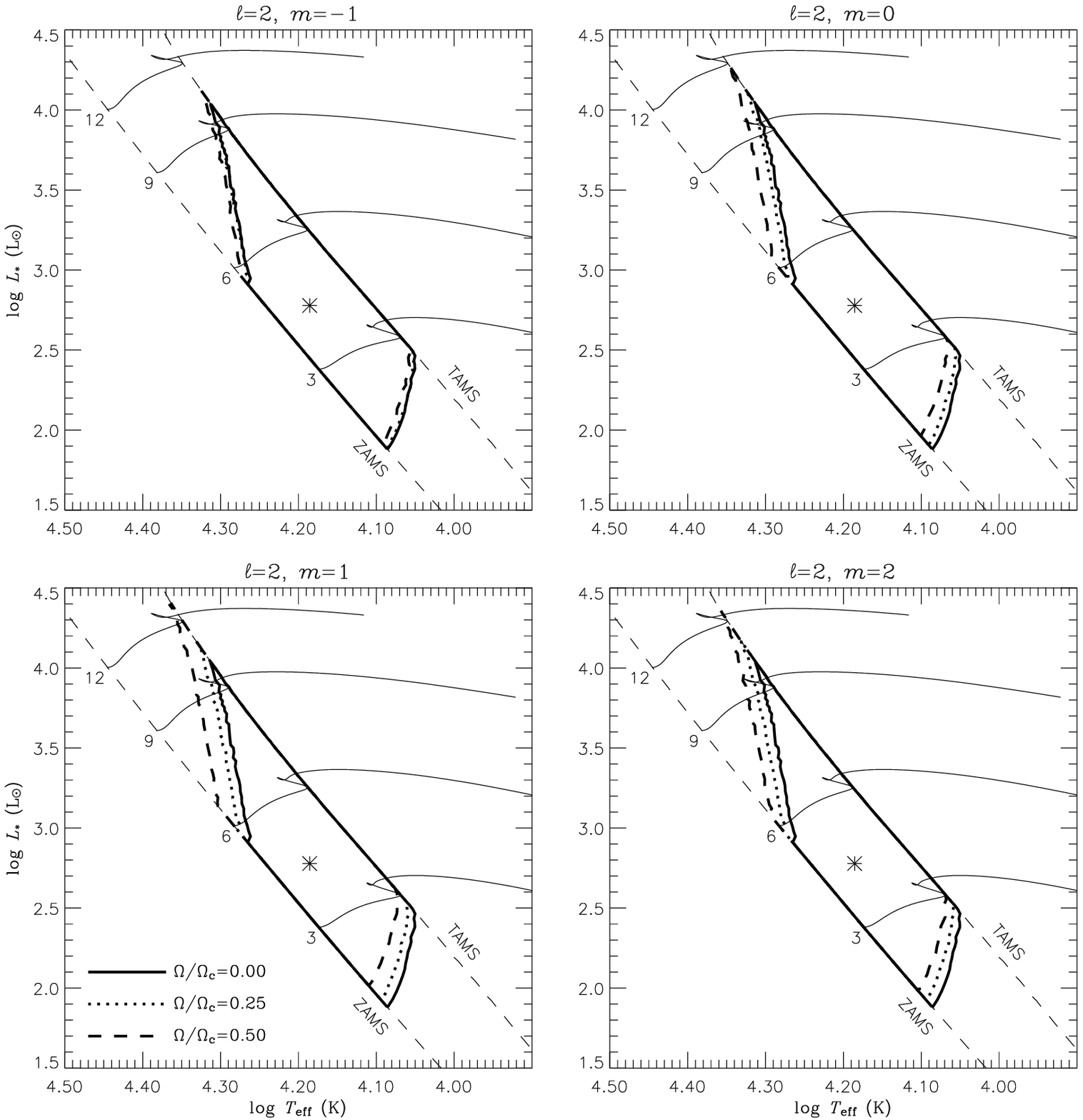}
\contcaption{}
\end{center}
\end{figure*}

The focus is now broadened, from the 53 Per-like model considered in
the preceding section, to the complete set of stellar models
introduced in Section~\ref{sec:models}. At three differing rotation
rates, $\rfreq/\rfreqc = 0.0$, 0.25 and 0.5, \boojum\ is used to
search for unstable $\ell=1$ and $\ell=2$ g modes; the results of
these calculations are presented in Fig.~\ref{fig:instab}, where the
same HR diagram shown in Fig.~\ref{fig:models} is overplotted by the
instability strip associated with each $(\ell,m)$ combination at each
rotation rate. These instability strips enclose all models that are
unstable ($\pfreqi < 0$) toward the excitation of one or more modes of
the indicated type.

As with the preceding section, there is a clear division between the
PS and the non-PS modes. The effect of rotation on the former is to
shift their instability strips along the main sequence, toward lower
temperatures and luminosities. The converse behaviour is exhibited by
the latter, for which instability strips are shifted toward higher
temperatures and luminosities. These shifts affect the blue edges of
the instability to a greater degree than the corresponding red
edges. For instance, the stellar models lying along the ZAMS are, in
the absence of rotation, unstable toward $(\ell,m)=(1,1)$ modes over
the temperature range $4.06 \leq \log \Teff \leq 4.21$. Upon the
introduction of rotation, at the rate $\rfreq/\rfreqc=0.5$, this range
shifts and broadens to $4.08 \leq \log \Teff \leq 4.28$.

\begin{figure*}
\epsffile{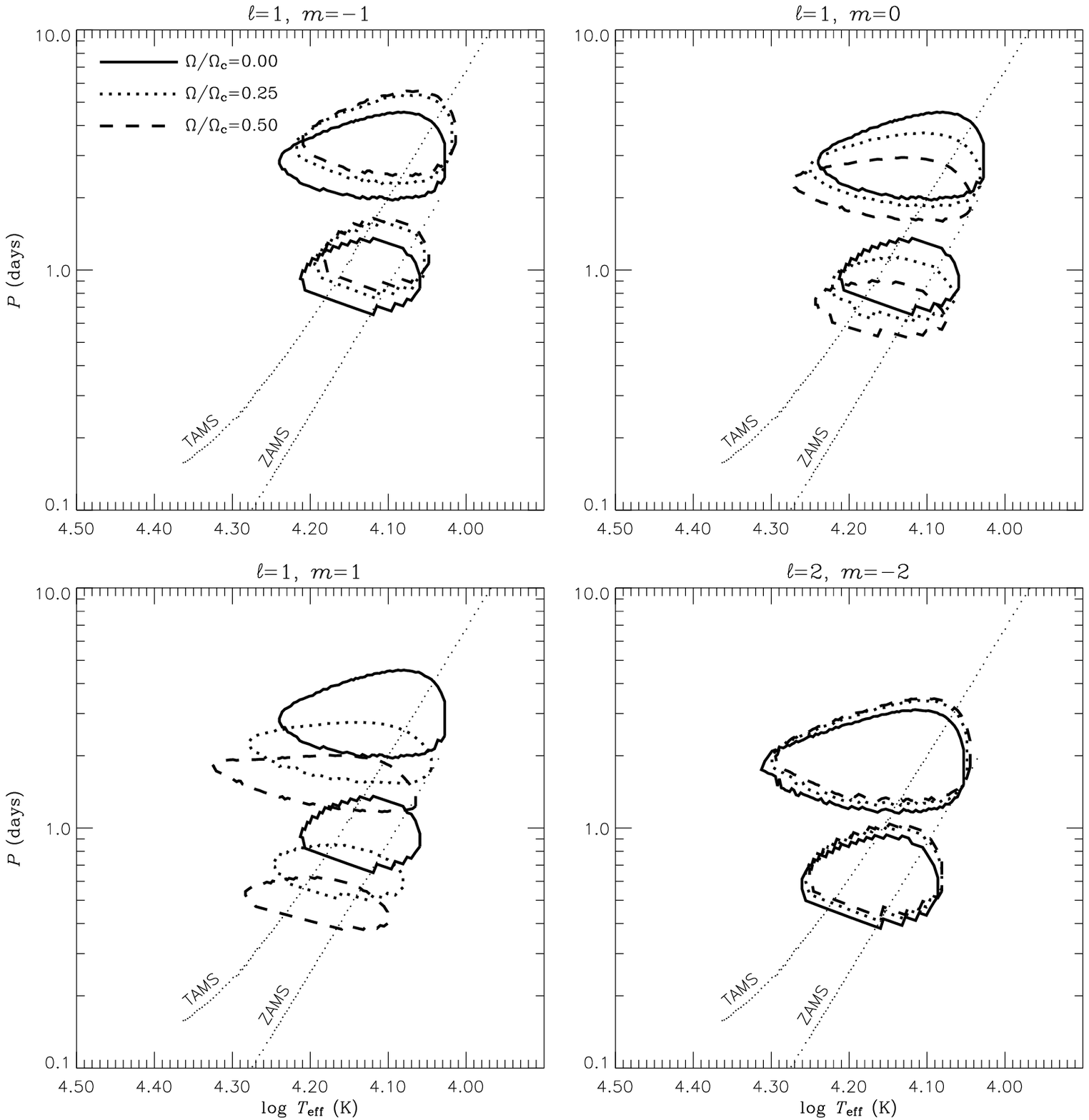}
\caption{Instability regions in the \Teff--\pper\ plane, for $\ell=1$
and $\ell=2$ g modes of stellar models lying along the ZAMS and TAMS
boundaries of the main sequence (\cf\ Fig.~\ref{fig:models}). In each
panel, corresponding to a particular combination $(\ell,m)$ of mode
parameters, the extent of the $\kappa$-mechanism instability is
indicated using thick lines, at the three differing rotation rates
$\rfreq/\rfreqc=0.0$ (solid), 0.25 (dotted) and 0.5 (dashed)
considered. The points, comprising separate curves for the ZAMS and
the TAMS models, indicate the thermal timescale \tauth\ within the
excitation zone.}
\label{fig:periods}
\end{figure*}

\begin{figure*}
\epsffile{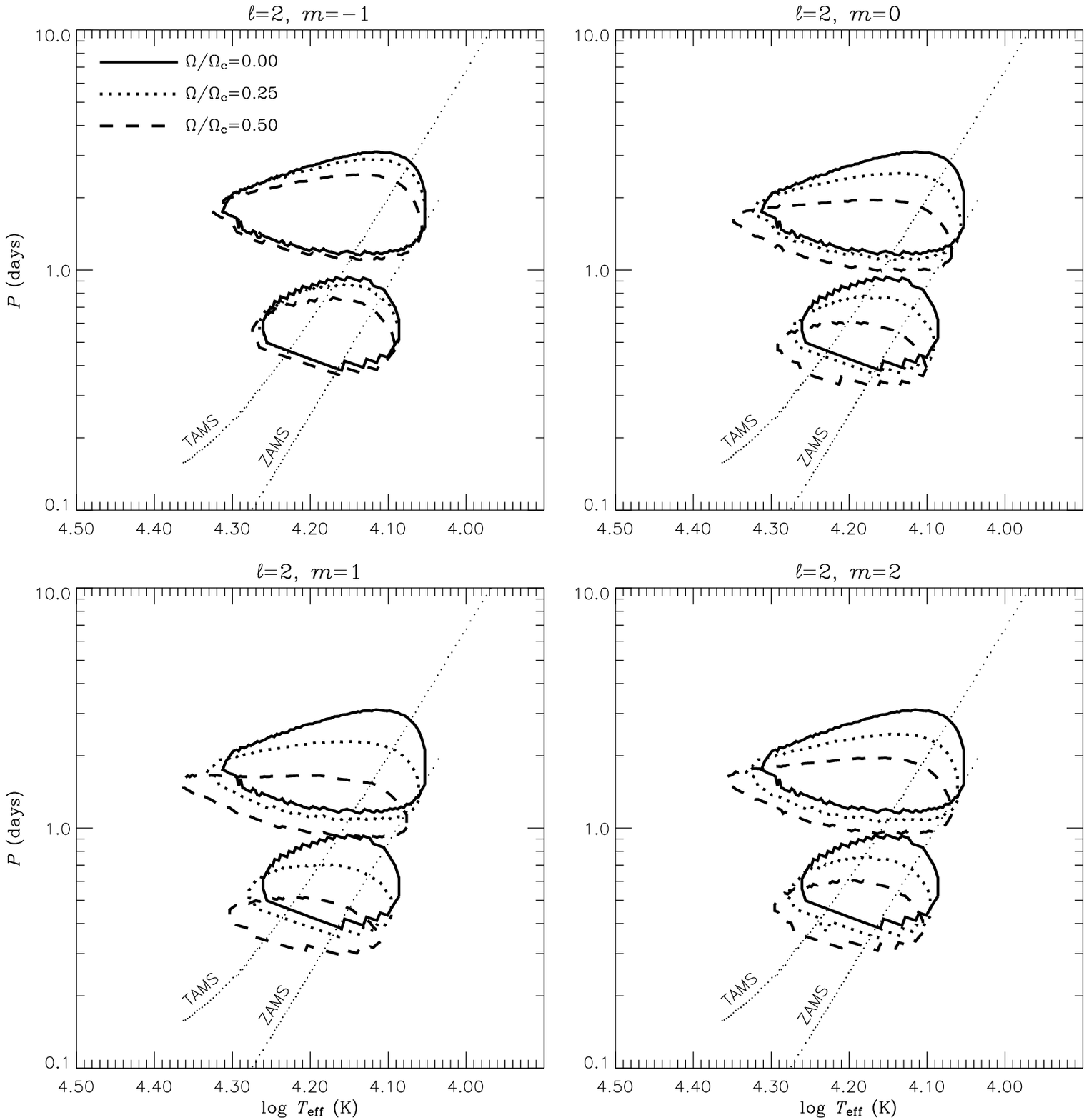}
\contcaption{}
\end{figure*}

The behaviour of \emph{both} classes of mode can be understood with
reference to the prerequisites for excitation discussed by
\citet{Dzi1993}. In addition to the aforementioned restriction on the
pressure eigenfunction (\cf\ Sec.~\ref{ssec:53per}), it is necessary
that the thermal timescale in the excitation zone be comparable to, or
longer than, the pulsation period. This latter condition is the key to
the behaviour seen in Fig.~\ref{fig:instab}. The thermal timescale
\begin{equation} \label{eqn:timescale}
\tauth(r) = \frac{\int_{r}^{\Rstar} T c_{p}\, \diff M_{r}}{\Lstar},
\end{equation}
evaluated at the radius where the opacity derivative $\kappa_{T}$ is
maximal, is a strong function of stellar effective temperature; it
declines rapidly toward larger values of \Teff, because the excitation
zone\footnote{Typically occurring at a temperature $\log
T\approx5.3$.} is situated ever closer to the surface. Consequently,
for the non-PS modes, whose periods become shorter under the influence
of the Coriolis force, the effect of rotation is to shift the
instability toward stars having higher \Teff, so that the
\tauth--\pper\ matching condition can still be met. Conversely, for
the PS modes, the marginal lengthening in periods due to rotation
pushes the instability toward slightly lower \Teff.

Fig.~\ref{fig:periods} illustrates these processes, by plotting
instability regions in the \Teff--\pper\ plane for stellar models
situated along the ZAMS and TAMS boundaries of the main
sequence. These regions, the complements of the instability strips
shown in Fig.~\ref{fig:instab}, indicate how the range of
unstable-mode periods evolves with changing effective temperature;
also shown in the figure, as separate curves for the ZAMS and the TAMS
models, is the thermal timescale within the excitation zone. For the
non-PS modes, the period shortening caused by rotation is evidently
responsible for the shift of the $\kappa$-mechanism instability toward
higher effective temperatures, in order to maintain the loose
correspondence between \tauth\ and \pper. Likewise, for the PS modes,
the period lengthening causes the instability to shift to lower
effective temperatures, once again to keep \tauth\ in step with \pper.

It is instructive to relate the foregoing analysis to previous studies
\citep[\eg, by][]{BalDzi1999} of the effect that a varying harmonic
degree $\ell$ has on the instability of SPB stars. It can be
recognized that the pulsation
equations~(\ref{eqn:puls-disp}--\ref{eqn:puls-enrg}) are -- modulo a
number of approximations -- no different than those describing modes
of harmonic degree \elle\ \emph{in a non-rotating star}. Accordingly,
the influence of the Coriolis force on the period and stability of an
individual mode can be followed simply by allowing $\elle$ to vary in
some specified manner. For non-PS modes, this variation assumes the
asymptotic form given in equation~(\ref{eqn:elle-asymp}); clearly, any
increase in $\nu$, corresponding to more-rapid rotation, leads to a
proportional increase in $\elle$. But, as the instability strips
presented by \citet[][their Fig.~5]{BalDzi1999} reveal, the result of
raising the harmonic degree is to shift the g-mode instability toward
higher luminosities and effective temperatures -- exactly the
behaviour manifested in the blueward displacement, at higher rotation
rates, of the instability strips plotted in Fig.~\ref{fig:instab}. A
similar line of reasoning can be applied to the PS modes.

Although these parallels to the non-rotating case are instructive, an
important caveat should be made regarding their use. Reiterating the
fact that the spin parameter $\nu$ is itself a function of the
pulsation frequency \pfreq, it is clear that the effective harmonic
degree $\elle$ assumes different values even among modes having the
same true harmonic degree $\ell$ and azimuthal order $m$. This result,
already discussed in Section~\ref{ssec:53per} under the guise of
spectral compression, is why the Coriolis force tends not only to
shift the $\kappa$-mechanism instability strips for SPB stars, but
also acts to broaden them.


\section{Discussion \& Summary} \label{sec:summary}

In the preceding sections, an approximate method for treatment of the
Coriolis force (Sec.~\ref{ssec:theory}) is used to devise equations
governing nonadiabatic, nonradial pulsation of rotating stars
(Secs.~\ref{ssec:equations}). These equations are solved
(Sec.~\ref{ssec:implement}) for a range of mid-B type stellar models
(Sec.~\ref{sec:models}); the general finding (Sec.~\ref{sec:calc}) is
that the Coriolis force shifts the instability strip associated with
$\kappa$-mechanism excitation of g modes, toward higher luminosities
and effective temperatures for non-PS modes, and toward lower
luminosities and effective temperatures for prograde sectoral modes.

An immediate corollary of this result is that the Coriolis force is
unable to stabilize \emph{all} B-type stars against g mode pulsation;
rather, it can only alter which particular stars -- at a given
rotation rate -- are unstable. Similar conclusions were reached by
\citet{UshBil1998} in their quasi-adiabatic analysis, and by
\citet{Lee2001} in his complete treatment of the Coriolis
force. However, as discussed in Section~\ref{sec:intro}, observations
of SPB stars reveal an apparent paucity of objects rotating at
significant rates. How might this discrepancy between theory and
observations be resolved?

One possibility is that the centrifugal force, neglected in the
present treatment and by \citet{Lee2001}, can act to stabilize
modes. Certainly, not taking this force into account will introduce a
certain level of error in the results -- especially at the upper limit
$\rfreq/\rfreq = 0.5$ adopted herein, which comes very close to
violating the assumption (\cf\ Sec.~\ref{ssec:theory}) that
$\rfreq^{2} \ll \rfreqc^{2}$. However, \citet{LeeBar1995} and
\citet{Lee1998} have demonstrated that the centrifugal force has
little effect on the stability of the g modes characteristic to SPB
stars. Perhaps a more plausible scenario is that resonant coupling,
between modes of differing harmonic degree \citep[see,
\eg][]{ChaLeb1962}, may act to suppress the $\kappa$-mechanism
instability. These couplings, which are neglected in the present
analysis because of the adoption of the traditional approximation,
were shown by \citet{Lee2001} to inhibit the excitation of selected
modes that would otherwise be unstable in rotating SPB stars.

Nevertheless, it is highly unlikely that resonant coupling could
suppress the instability of \emph{all} modes. With this in mind, it
appears increasingly probable that, as the theoretical analysis
indicates, g modes \emph{are} excited in rotating SPB stars; but that
the equatorial-confinement effects demonstrated by \citet{Tow2003b}
render them very difficult to detect observationally. In support of
this interpretation for the observational scarcity of rotating SPB
stars, two of the open clusters originally thought to be devoid of SPB
stars -- NGC~4755 \citep{BalKoe1994} and NGC~6231 \citep{BalLan1995}
-- have, upon closer scrutiny, been found to contain a number of
candidate-SPB objects \citep[see][]{Sta2002,Are2001}.

In addition to their direct applicability to SPB stars, the results
presented herein are highly relevant to the understanding of the Be
phenomenon. The multiperiodic spectroscopic and photometric variations
exhibited by many Be stars (\eg, $\mu$ Cen -- \citealt{Riv1998};
$\omega$ CMa -- \citealt{Ste2003}) are usually interpreted as arising
from nonradial g modes \citep[see][and references
therein]{Riv2003}. It is not unreasonable to suppose the excitation of
these modes to be due to the same $\kappa$ mechanism operative in the
SPB stars. However, a historical problem with this stance has been
that many variable Be stars have early spectral types B0--B3, falling
\emph{blueward} of the high-temperature limit of the SPB instability
strip for low-degree pulsation \citep{BalDzi1999}. Furthermore,
periodic variations are not seen in Be stars with types later than B5
\citep[\eg,][and references therein]{Baa1989a,Baa1989b}, even though
the SPB instability strip extends all the way down to types B8 or B9.

Both of these difficulties are resolved by allowing for the influence
of the Coriolis force, which -- as the present analysis demonstrates
-- has the effect of shifting $\kappa$-mechanism instability toward
earlier spectral types. Of course, this rationalization can only work
if the modes seen in Be stars are \emph{not} prograde sectoral;
however, in light of the analysis presented by \citet{Riv2003}, who
found at least 16 of the 27 Be stars in their sample to pulsate in
\emph{retrograde} $(\ell,m) = (2,2)$ modes, this latter restriction
does not appear to be problematic.


\section*{Acknowledgments}

I thank Stan Owocki and Myron Smith for their helpful suggestions, and
the referee Prof. W. Dziembowski for his insightful remarks. This
research has been partially supported by US NSF grant AST-0097983, and
by the UK Particle Physics and Astronomy Research
Council. Computations were undertaken on the Parameter Search Engine
at the HiPerSPACE Computing Centre, UCL, which is funded by PPARC.


\bibliography{coriolis}

\bibliographystyle{mn2e}


\label{lastpage}

\end{document}